\documentclass[journal=jpclcd,manuscript=letter]{achemso}
\usepackage{chemformula} 
\usepackage[T1]{fontenc}

\usepackage{mathptmx}
\usepackage{amsmath,amssymb}
\usepackage{graphicx}
\usepackage{braket}
\usepackage{setspace}
\usepackage{xcolor}
\usepackage[normalem]{ulem}
\usepackage{float}   

\AtBeginDocument{\def\hbar{\bar{h}}}

\newcommand{\Pload}{P_{\mathrm{load}}}
\newcommand{\Psel}{P_{\mathrm{sel}}}
\newcommand{\Pad}{P_{\mathrm{ad}}}
\newcommand{\Pmap}{P_{\mathrm{map}}}
\newcommand{\Psurv}{P_{\mathrm{surv}}}

\author{Meng-Yi Yu}
\affiliation[1st University]
{Hefei National Research Center for Physical Sciences at the Microscale, University of Science and Technology of China, Hefei 230026, China}

\author{Ya-Nan Lv}
\email{yananlv@ustc.edu.cn}
\affiliation[1st University]
{Hefei National Research Center for Physical Sciences at the Microscale, University of Science and Technology of China, Hefei 230026, China}
\alsoaffiliation[2nd University]
{Hefei National Laboratory, University of Science and Technology of China, Hefei 230088, China}

\author{Cun-Feng Cheng}
\email{cfcheng@ustc.edu.cn}
\affiliation[1st University]
{Hefei National Research Center for Physical Sciences at the Microscale, University of Science and Technology of China, Hefei 230026, China}
\alsoaffiliation[2nd University]
{Hefei National Laboratory, University of Science and Technology of China, Hefei 230088, China}

\author{Shui-Ming Hu}
\affiliation[1st University]
{Hefei National Research Center for Physical Sciences at the Microscale, University of Science and Technology of China, Hefei 230026, China}
\alsoaffiliation[2nd University]
{Hefei National Laboratory, University of Science and Technology of China, Hefei 230088, China}
\alsoaffiliation[3rd University]
{State Key Laboratory of Chemical Reaction Dynamics, University of Science and Technology of China, Hefei 230026, China}

\title
  {Detuning-Controlled Vibrational Autler--Townes Passage toward Single-Sublevel Orientation of Nonpolar Molecules}

\abbreviations{AT, ASAP, OPO, FWHM}

\setkeys{acs}{articletitle=true}

\begin{document}
\makeatletter
\setlength\acs@tocentry@height{1.75in}
\setlength\acs@tocentry@width{3.25in}
\makeatother
\begin{tocentry}
  \centering
  \includegraphics[width=3.2in,height=1.72in,keepaspectratio]{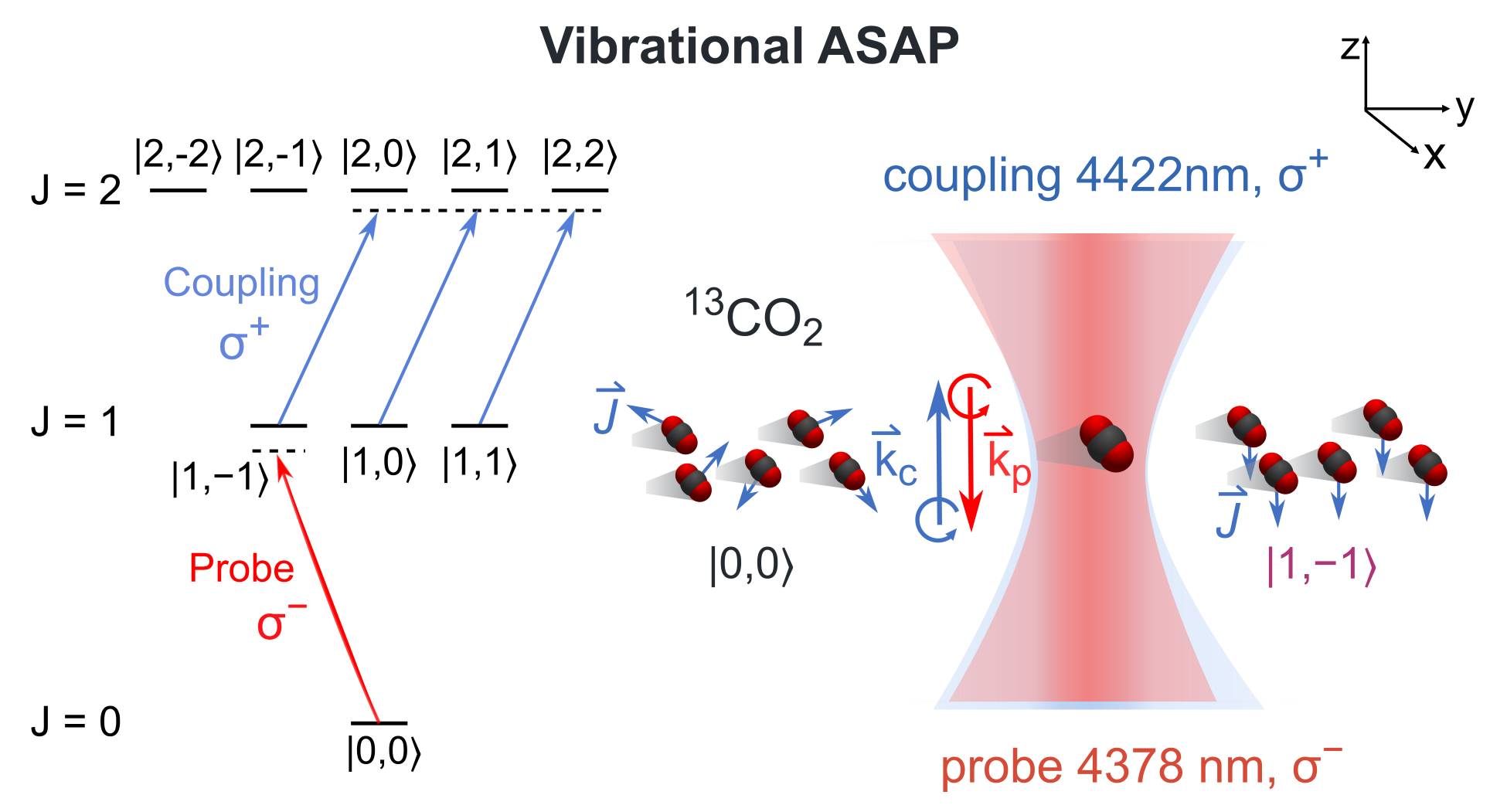}
\end{tocentry}
\begin{abstract}
Single-sublevel preparation would give collision and surface-scattering experiments access to oriented molecules, yet established orientation methods either require a permanent dipole moment or select only partially. We introduce all-optical Autler--Townes state-selective adiabatic passage (ASAP), in which a detuned infrared field dresses a rovibrational transition, and the selected dressed branch is mapped adiabatically onto one magnetic sublevel as the field vanishes. We resolve the magnetic-sublevel-selected rovibrational Autler--Townes doublet of the R(0) transition of $^{13}$CO$_2$ and show that the coupling strength and detuning 
set both its separation and the partition of the bare-state composition between its two components. 
The spectral selectivity of the target dressed branch reaches 97\% at a coupling detuning of $-1.2$~MHz. 
A density-matrix model calibrated against these spectra returns, for the dressed molecules, a target-sublevel population approaching 0.9 over a broad range of coupling powers and detunings.
\end{abstract}



The outcome of a molecular encounter depends on how the partners are turned toward each other. In a thermal ensemble the orientational average washes this dependence out, so that extracting molecular-frame information from collision, scattering, or reaction experiments requires an ensemble whose orientational distribution is known, ideally one in which every molecule occupies a single angular-momentum sublevel along a chosen laboratory axis. We use oriented in precisely this sense throughout: one magnetic sublevel of the angular momentum relative to the laboratory quantization axis, as distinct from a static orientation of the molecular axis or of a permanent dipole moment. With such an ensemble the steric effect becomes something the experimenter can vary~\cite{lemeshko2013,koch2019,brooks1976,polanyi1987,vallance2011,yoder2010,hou1997}, and it is also the prerequisite for preparing single-quantum-state reactants in cold-collision and quantum-information studies~\cite{Carr2009NJP,Krems2008PCCP,Ni2008Science,DeMille2002PRL}.

Any realistic implementation has to satisfy several conditions at once: the magnetic sublevel degeneracy must be lifted and resolved; population must be transferred into one sublevel rather than distributed over several; the prepared orientation must survive long enough to be transported to the interaction region; and, for collision experiments specifically, the molecules must afterwards be free of external fields, since any residual electric or magnetic field would itself steer the encounter being studied.

Established approaches meet these requirements only in part, and each is confined to a restricted class of molecules. Electrostatic techniques such as hexapole focusing and brute-force static fields do select individual sublevels, but operate only on species possessing a permanent dipole moment~\cite{kramer1965,cho1991,loison1995,Friedrich1991Nature,Goban2008PRL,Ghafur2009NP}. Intense nonresonant laser pulses create alignment rather than orientation, because the interaction depends on the square of the field amplitude and thus produces head-versus-tail symmetry~\cite{kim1996,sakai1999,larsen1999,stapelfeldt2003}. Supersonic expansions and resonant optical pumping produce only partial selectivity, and only for favorable level structures~\cite{pullman1990,aquilanti1994,pirani2001,scoles1988}.

Dressing a transition with a strong infrared field removes two of these obstacles in a single step: it resolves the sublevels and it supplies the mechanism that moves population into the selected one~\cite{spano2001,stahovich2024}. The two transitions share a common level, which is the upper level of the probed transition and the lower level of the coupling transition. When a strong coupling field drives that common level, the probe resonance splits into two components whose separation and relative intensities are controlled by the field. In the dressed-state description~\cite{CohenTannoudjiHaroche1969,Mollow1969} the coupling field mixes the two bare states it couples into two eigenstates separated in energy by $\hbar R_{M}$, where
\begin{equation}
R_M=\sqrt{\Omega_{c,M'}^2+\Delta_c^2}
\label{eq:Rabi}
\end{equation}
is the generalized Rabi frequency, $\Omega_{c,M'}$ measures the coupling strength set by the transition dipole moment and the field amplitude, and $\Delta_c$ is the detuning of the coupling laser from resonance. Numbers quoted below and in the figures are $R_M/2\pi$, $\Omega_{c,M'}/2\pi$ and $\Delta_c/2\pi$, that is, hertz rather than radians per second. The two components therefore carry complementary fractions of the two bare states, and those fractions are set by the mixing angle $\theta_M$ through $\cos^2\theta_M=(R_M+\Delta_c)/2R_M$.

The sublevels become distinguishable because a circularly polarized coupling field addresses each $M$ channel through its own Clebsch--Gordan factor, so different channels acquire different effective coupling strengths and hence different splittings $R_M$. Where several channels are driven simultaneously they appear at different probe frequencies and can be addressed one at a time. This route was recently validated in Li$_2$ using dressed transitions to an electronically excited level~\cite{stahovich2024}. That demonstration, however, inherits the nanosecond radiative lifetime of an electronic excited state, too short for the prepared molecules to reach a collision region, and the dressed states it prepares are superpositions of two bare states, so only part of the population reaches a single sublevel.

The scheme presented here, vibrational Autler--Townes state-selective adiabatic passage (ASAP), addresses both limitations. First, the dressing is applied to vibrational transitions within the electronic ground state, for which radiative lifetimes are of order milliseconds, $\tau\approx2.7$~ms for the level used below~\cite{GORDON2026HITRAN2024}, rather than nanoseconds. Second, the coupling-laser detuning makes the composition of the dressed states something one can set, and it is this that the scheme rests on. For $\Delta_c=0$ the mixing angle is fixed at $\pi/4$ and each dressed branch contains half of the target bare state, so even perfect transfer into one branch leaves no more than half of the population in that state. For $\Delta_c<0$ the selected branch carries the larger share of the target state, and because $R_M\to|\Delta_c|$ while $\Omega_{c,M'}\to0$ at the trailing edge of the field, that share tends to unity as the molecules leave the beams. Provided the dressed-state composition changes slowly compared with $R_M^{-1}$, the population follows its branch adiabatically and emerges in a single bare sublevel. Detuning is therefore not a refinement of the resonant scheme but what turns dressed-state preparation into, in principle, a nearly complete population transfer. Because the fields vanish at the end of the interaction, the molecules leave as freely flying oriented particles with no field to perturb the experiment that follows.

\begin{figure}[H]
    \centering
    \includegraphics[width=0.8\textwidth]{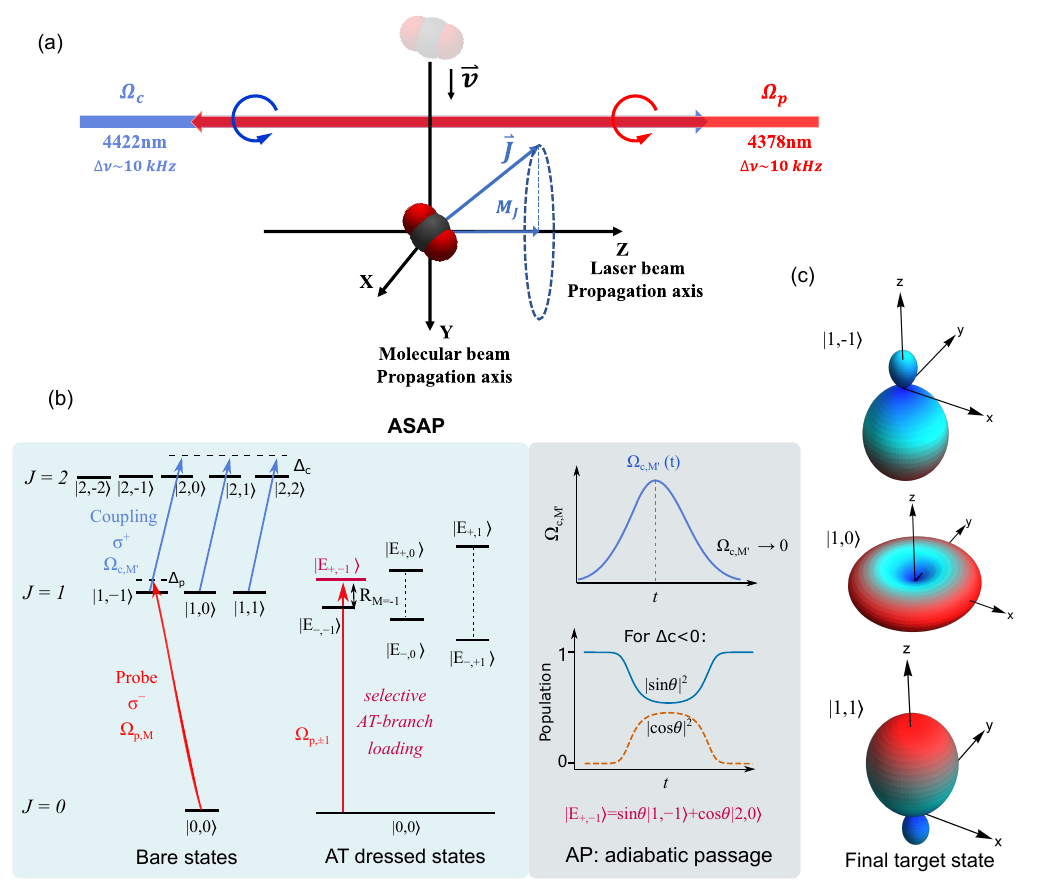}
    \caption{Schematic of the experimental geometry and the ASAP protocol. (a) A molecular beam along $+y$ crosses counter-propagating $\sigma^{-}$-polarized probe and $\sigma^{+}$-polarized coupling beams along the $z$ quantization axis. (b) The probe and coupling fields connect $\ket{0,0}$, $\ket{1,M}$, and $\ket{2,M'}$ in three vibrational levels, whose vibrational labels are omitted. The coupling field mixes $\ket{1,M}$ and $\ket{2,M'}$ into dressed states, and the probe frequency selects the target $\ket{E_{+,M}}$ branch. The illustrated pathway corresponds to $M=-1$ and $M'=0$ for the polarizations in (a). The upper right curve shows the Gaussian coupling envelope experienced by the molecule. For $\Delta_c<0$, the lower curves show the $\ket{1,M}$ and $\ket{2,M'}$ weights $\sin^2\theta_M$ and $\cos^2\theta_M$ of $\ket{E_{+,M}}$, illustrating adiabatic mapping onto $\ket{1,M}$. Here $R_M$ denotes the dressed-branch separation in angular-frequency units. (c) Selectable $J=1$ sublevels ($M=-1,0,+1$) and the corresponding spatial distributions of the angular momentum along $z$, accessible via appropriate polarization and beam geometries.}
    \label{Fig:setup}
\end{figure}

What follows establishes the spectroscopic prerequisite of ASAP and uses a calibrated model to quantify the achievable transfer. 
The experiments are carried out on $^{13}$CO$_2$, a nonpolar molecule that established orientation methods cannot address. As shown in Fig.~\ref{Fig:setup}(a), a pulsed molecular beam crosses two counter-propagating, circularly polarized light fields along the quantization axis, linking the $J=0$, 1, 2 ladder of Fig.~\ref{Fig:setup}(b). The weak $\sigma^{-}$ probe couples $\ket{0,0}\to\ket{1,M}$ at Rabi frequency $\Omega_{p,M}$ and detuning $\Delta_{p}$, while a strong $\sigma^{+}$ coupling field drives $\ket{1,M}\to\ket{2,M'}$ at $\Omega_{c,M'}$ and $\Delta_{c}$. The detunings are defined as $\Delta_p=\omega_{10}-\omega_p$ and $\Delta_c=\omega_{21}-\omega_c$, where $\omega_{10}$ and $\omega_{21}$ are the bare transition angular frequencies of $\ket{0,0}\to\ket{1,M}$ and $\ket{1,M}\to\ket{2,M'}$ and $\omega_p$, $\omega_c$ are the laser frequencies. Under the weak-probe approximation the coupling field mixes each pair of bare states into two dressed states [Eq.~\eqref{eq:dressed}],
\begin{align}
    \ket{E_{-,M}}&= \cos\theta_M\ket{1,M}-\sin\theta_M\ket{2,M'}, \nonumber\\
    \ket{E_{+,M}}&=\sin\theta_M\ket{1,M}+\cos\theta_M\ket{2,M'},
    \label{eq:dressed}
\end{align}
split by $\hbar R_M$, with $\theta_M=\arccos\sqrt{(R_M+\Delta_c)/2R_M}$ controlled by $\Omega_{c,M'}$ and $\Delta_c$. Because the probe couples only to the $\ket{1,M}$ component, it excites either dressed state with effective strengths $\Omega_{p,M}\cos\theta_M$ and $\Omega_{p,M}\sin\theta_M$ for the $E_{-,M}$ and $E_{+,M}$ branches, respectively. The Clebsch--Gordan coefficients make $\Omega_{c,M'}$ sublevel dependent, so the branch energies vary with $M$ and a single branch, and with it a single sublevel, can be selected by tuning the probe frequency. The two resonances appear at probe detunings
\begin{equation}
\delta_{\mathrm{AT},\pm}=-\frac{\Delta_c}{2}\mp\frac{1}{2}
\sqrt{\Delta_c^2+\Omega_{c,M'}^2}.
\label{eq:peakpos}
\end{equation}
As molecules cross the Gaussian beams, adiabatic following requires $|\dot{\theta}_M(t)|/R_M(t)\ll1$ throughout the transit~\cite{Bergmann1998}. For $\Delta_c<0$, the selected $\ket{E_{+,M}}$ branch then maps onto $\ket{1,M}$ as $\Omega_{c,M'}(t)$ vanishes [Fig.~\ref{Fig:setup}(b)]. Reversing both laser helicities gives the corresponding pathway to $\ket{1,1}$, while $\pi$-polarized excitation in an appropriate beam geometry reaches $\ket{1,0}$, so every sublevel of Fig.~\ref{Fig:setup}(c) becomes accessible.

$^{13}$CO$_2$ molecules seeded in argon (3\%) are expanded in a pulsed supersonic beam through an Even-Lavie valve and collimated by a skimmer. The beam has a central axial velocity of $\approx$700~m/s with a spread of $\approx\pm$40~m/s ($3\sigma$) and a rotational temperature of $\approx$8~K. Coupling (4422~nm) and probe (4378~nm) radiation is generated by continuous-wave optical parametric oscillators (OPOs) frequency-stabilized at the 10~kHz level~\cite{zhangOE2020,Tan2024CJCP-OPO,Vainio2016PCCP}. The source architecture is described in Section S1 of the Supporting Information. Both beams are circularly polarized by independent quarter-wave plates and focused to a diameter of 1~mm, giving a transit time of $\approx$1.4~$\mu$s. The coupling beam delivers up to 60~mW, while the probe power is held below 10~$\mu$W so that the weak-field approximation remains valid and additional broadening is avoided. The coupling laser drives the R-branch transition $J=1$ $(000^11)\to J=2$ $(000^21)$ and the probe interrogates $J=0$ $(000^01)\to J=1$ $(000^11)$, in the standard $(n_1n_2n_3lr)$ notation~\cite{Herzberg1945}. The R-branch choice ensures that all three $M$ sublevels of $J=1$ have non-vanishing transition dipoles to $J=2$, so the circularly polarized coupling field addresses every sublevel with $M$-dependent Rabi frequencies. We take $\ket{1,-1}$ as the target state. By adjusting $\Omega_{c,M'}$, $\Delta_c$, $\Omega_{p,M}$, and the laser polarizations, the same apparatus either resolves the $E_{\pm,M}$ branches, as in the measurements below, or prepares the target state [Fig.~\ref{Fig:setup}(c)].

\begin{figure}[H]
    \centering
    \includegraphics[width=0.8\textwidth]{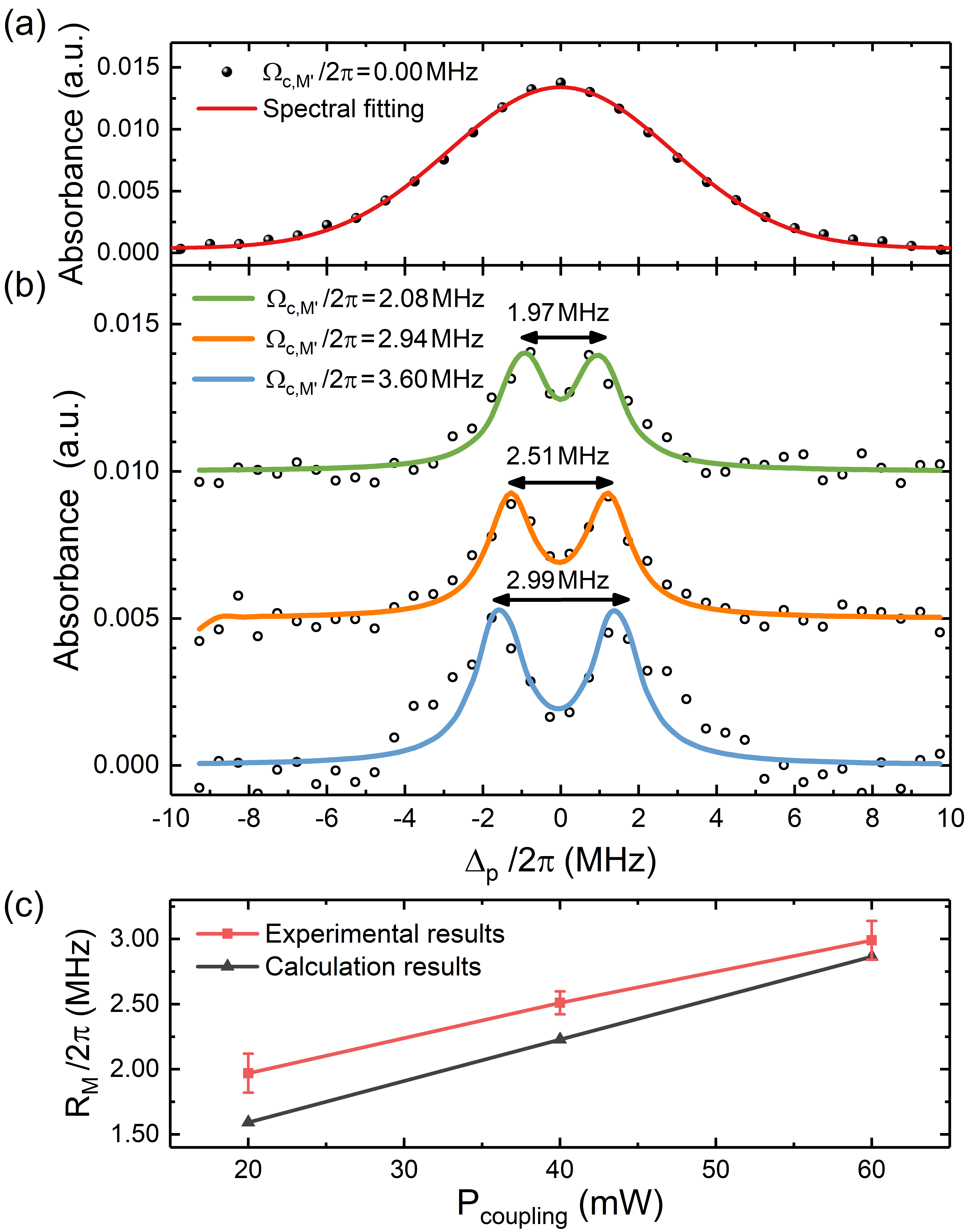}
    \caption{Vibrational Autler--Townes splitting of the $^{13}$CO$_2$ R(0) rovibrational transition. (a) Probe absorption spectrum without the coupling field. The black circles are experimental data and the red line is the spectral fit. (b) Spectra after subtraction of the fitted Doppler background at three coupling Rabi frequencies $\Omega_{c,M'}/2\pi$, vertically offset for clarity. Open circles are experimental data and solid lines are fitted spectra. (c) Measured dressed-branch separation $R_{M=-1}$ versus coupling laser power. Red squares with error bars are experimental results, black triangles simulations. In (b) and (c) the coupling field is resonant with the $\ket{1,-1}\to\ket{2,0}$ transition, i.e. $M=-1$, $M'=0$. }
    \label{Fig:ATspec}
\end{figure}

Probe transmission was recorded while the probe frequency was scanned stepwise across the resonance, averaging over successive gas pulses. Without the coupling field, the absorption shows the expected supersonic-beam Doppler profile with a full width at half maximum (FWHM) of 6.77~MHz [Fig.~\ref{Fig:ATspec}(a)]. With the coupling laser on, a pair of dressed-state absorption features appears on the residual Doppler background and separates systematically as the coupling intensity increases [Fig.~\ref{Fig:ATspec}(b)]. Subtracting the fitted Doppler contribution isolates cleanly resolved components (Section S2 of the Supporting Information). 
The doublet identified above belongs to a single magnetic channel, and its isolation from the remaining channels rests on two independent mechanisms. With the polarizations used here, the sequential selection rules alone restrict the spectrum to one channel: acting on the unique $M=0$ ground state, the $\sigma^-$ probe populates only $M=-1$ of $J=1$ by the $\Delta M=-1$ rule, and only this sublevel is then addressed by the $\sigma^+$ coupling through the $M=-1\to0$ channel. Independently, the $M$ dependence of $R_M$ described above would separate any further channels in probe frequency, so the protocol does not rely on the polarization selection rules alone. Interchanging the helicities addresses $M=+1$ symmetrically, so all sublevels are reachable in the same apparatus. The measured splitting $R_{M=-1}$ grows with coupling power [Fig.~\ref{Fig:ATspec}(c)] and agrees quantitatively with density-matrix simulations at experimentally calibrated power densities (Section S3 of the Supporting Information).

\begin{figure}[H]
    \centering
    \includegraphics[width=0.8\textwidth]{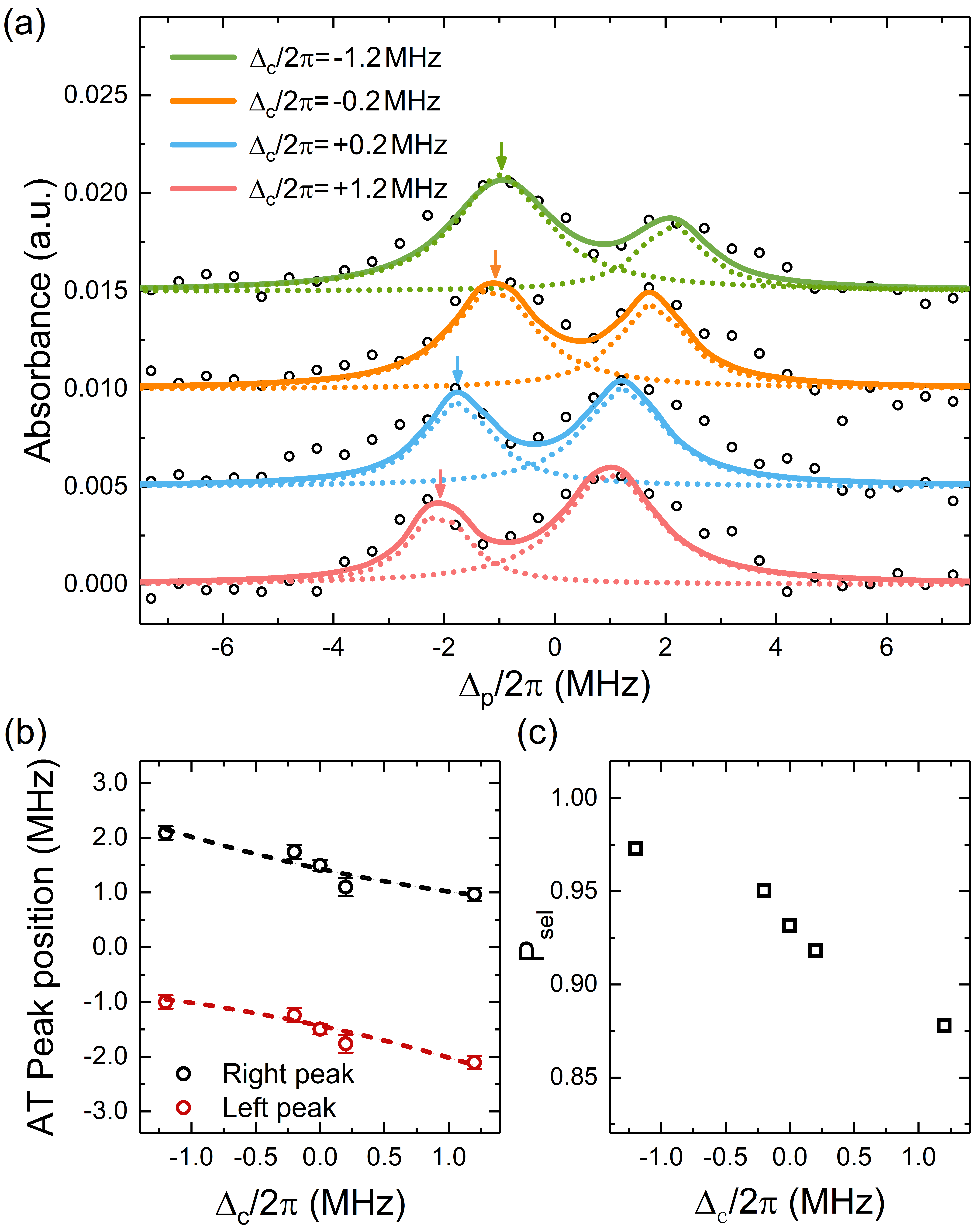}
    \caption{Coupling-detuning dependence of the rovibrational doublet and of the branch selectivity. (a) Spectra at fixed coupling power and different coupling detunings, vertically offset for clarity. Open circles are experimental data, solid lines spectral fits, and dotted curves the fitted contributions of the two branches. Arrows mark the fitted centers of the selected $E_{+,M}$ branch, which fix the excitation frequencies used in (c). (b) Fitted branch centers versus coupling detuning. The red and black circles denote the left and right components, error bars are one-standard-deviation uncertainties from the fit covariance matrix, and dashed lines are the positions from Eq.~\eqref{eq:peakpos}. (c) Spectral branch selectivity $P_{\mathrm{sel}}$ for the arrow-marked branch in (a) as a function of the coupling detuning $\Delta_c/2\pi$. The open squares give the selected branch's fractional contribution to the total fitted response at the marked excitation frequencies.}
    \label{Fig:ATDetuning}
\end{figure}

The composition of the doublet responds to the coupling detuning, and this is the degree of freedom the scheme relies on (Fig.~\ref{Fig:ATDetuning}). A nonzero $\Delta_c$ produces an asymmetric doublet: the component lying closer to $\Delta_p=0$ gains intensity because it carries the larger $\ket{1,M}$ weight [Fig.~\ref{Fig:ATDetuning}(a)], whereas at $\Delta_c=0$ the two components have equal strength. The fitted peak positions follow Eq.~\eqref{eq:peakpos} over the full range of detunings [Fig.~\ref{Fig:ATDetuning}(b)]. Because the two components correspond to complementary weights of $\ket{1,M}$ and $\ket{2,M'}$, their integrated strengths directly report how much of each dressed branch belongs to the target bare state, and hence how much population can in principle be mapped onto it.

Quantifying that selectivity is a prerequisite for using the doublet for state preparation, since any absorption from the unwanted branch populates the wrong dressed state. At each detuning we evaluate the selectivity of the target branch, $P_{\mathrm{sel}} =A_L(\Delta_{p,0})/[A_L(\Delta_{p,0})+A_R(\Delta_{p,0})]$, at its fitted peak center $\Delta_{p,0}$, where $A_L$ and $A_R$ are the fitted left and right contributions including amplitudes but excluding background [Fig.~\ref{Fig:ATDetuning}(c)]. The selectivity $P_{\mathrm{sel}}$ rises with $|\Delta_c|$ and reaches $P_{\mathrm{sel}}\simeq0.97$, the fit uncertainty being reported in Section~S5 of the Supporting Information, at $\Delta_c/2\pi=-1.2$~MHz. Because the measurement is made in the linear-absorption regime, this spectral ratio equals the ratio of populations loaded into the two branches by a weak resonant probe. This number therefore fixes the largest fraction of the excited population that can reach the target sublevel, reached only if the population loaded into the selected branch is subsequently mapped onto it perfectly.

\begin{figure}[H]
    \centering
    \includegraphics[width=0.8\textwidth]{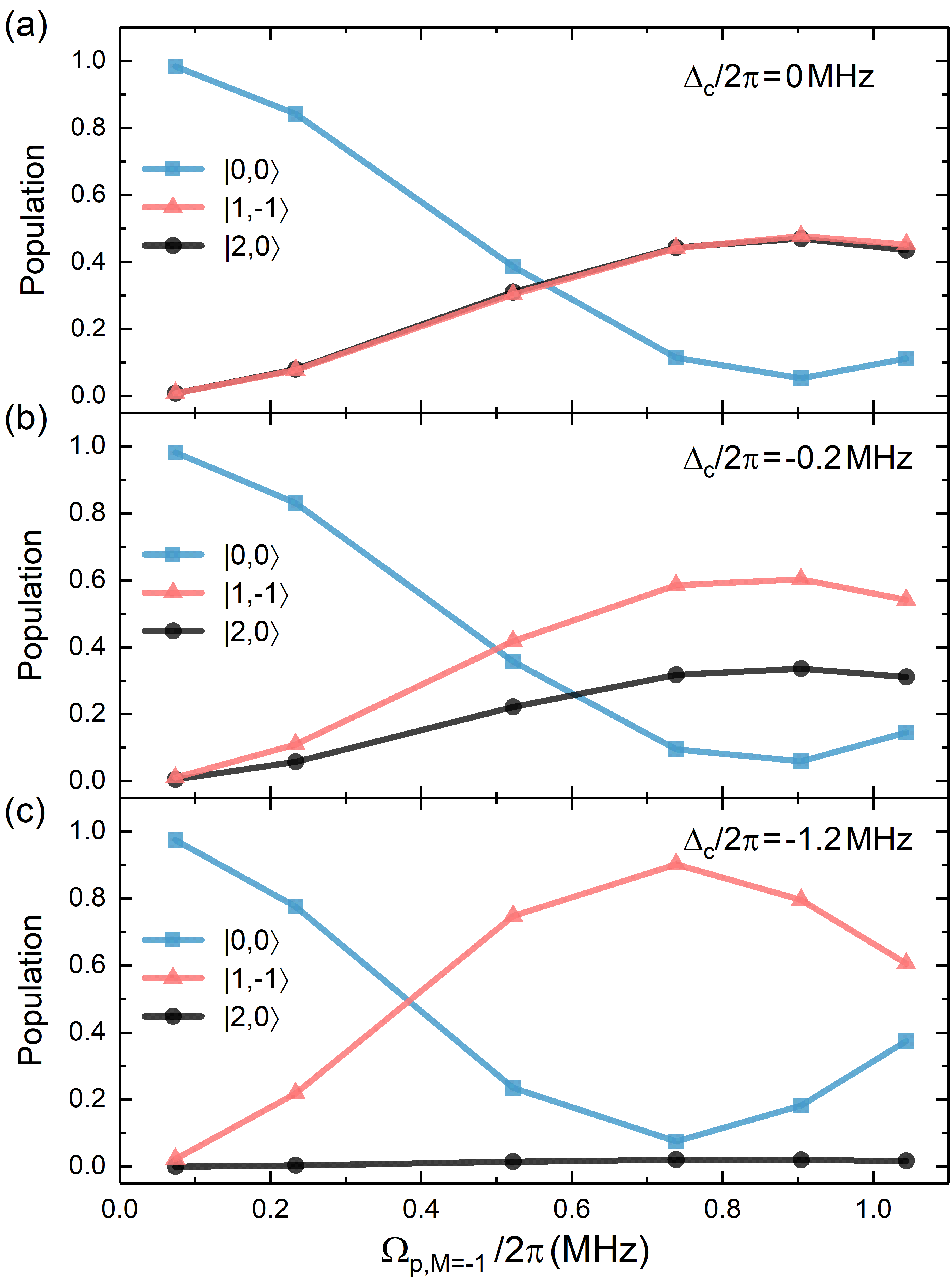}
    \caption{Simulated ensemble-averaged final bare-state populations of $\ket{0,0}$ (blue squares), the target state $\ket{1,-1}$ (red triangles), and $\ket{2,0}$ (black circles) versus the peak probe Rabi frequency $\Omega_{p,M=-1}/2\pi$, at coupling detunings (a) $\Delta_c/2\pi=0$~MHz, (b) $-0.2$~MHz, and (c) $-1.2$~MHz. The initial state is $\ket{0,0}$ and populations are evaluated after both fields have vanished. For each detuning the probe frequency is fixed at the target $E_{+,M}$ peak identified in the weak-probe limit and held constant throughout the power scan. Simulations use parameters matching the corresponding experimental conditions, except for the scanned probe power. The target-state population approaches 0.9 in (c).}
    \label{fig:signal_population}
\end{figure}

The final state population follows from a density-matrix model whose parameters are fixed by the spectra above, with the provenance of each parameter listed in Table~S2 of the Supporting Information, and which is averaged over the Gaussian laser profiles and over the velocity distribution of the molecules the coupling field dresses rather than over that of the beam as a whole (Sections S3 and S6 of the Supporting Information). Figure~\ref{fig:signal_population} shows the ensemble-averaged populations evaluated after both fields have vanished, as a function of probe power at three coupling detunings. Throughout, the quantity of interest is the final population $P_{\mathrm{final}}=\langle 1,-1|\bar{\rho}(t_f)|1,-1\rangle$, where $\bar{\rho}$ is the velocity-averaged density matrix and $t_f$ is a time after the molecules have left the fields. Since only the probe power is changed, the branch separation and the relative spectral weights of Fig.~\ref{Fig:ATDetuning} carry over as long as the probe leaves the dressed-state structure unperturbed. Two requirements then compete. The effective pulse area $\Theta$ determines how much population leaves the ground state, with the first maximum near $\Theta=\pi$, while the adiabatic-following probability and the final target-state weight of the selected branch determine how much of that population ends in $\ket{1,-1}$.

At resonance [Fig.~\ref{fig:signal_population}(a)] the mixing angle of the addressed channel stays at $\pi/4$ at all times, each dressed branch keeping half of $\ket{1,-1}$ and half of $\ket{2,0}$. Even perfect following of one branch therefore leaves its target weight at one half, so although the total excitation probability reaches approximately 0.94, the final population divides nearly equally between the two bare states and $P_{\mathrm{final}}\simeq0.46$. At $\Delta_c/2\pi=-0.2$~MHz the branch separation $\hbar R_M(t)$ on the trailing edge is still too small to maintain $|\dot{\theta}_M|/R_M\ll1$. Assuming ideal loading, the calculated adiabatic-following probability is approximately 0.66, so nonadiabatic transfer between the branches limits the outcome and $P_{\mathrm{final}}$ reaches only about 0.60 despite a similar excitation probability [Fig.~\ref{fig:signal_population}(b)]. Increasing the detuning to $-1.2$~MHz opens the gap during turnoff and raises the adiabatic-following probability to 0.99 [Fig.~\ref{fig:signal_population}(c)]. Near the optimum probe power the residual $\ket{2,0}$ population falls to about 0.02 and $P_{\mathrm{final}}$ approaches 0.9. To identify which stage limits the transfer, we factorize the result as $P_{\mathrm{final}}\simeq\Pload\,\Psel\,\Pad\,\Pmap\,\Psurv$, representing branch-selective loading, spectral selection, adiabatic following, dressed-to-bare mapping, and survival against spontaneous emission (Section S5 of the Supporting Information). Losses here are dominated by incomplete loading: raising the probe power beyond the optimum produces coherent overrotation that lowers both the excitation probability and the target population. High efficiency persists over a wide range of coupling powers and detunings and is relatively insensitive to the longitudinal velocity spread, while the transverse velocity width is the main limitation, because it determines how large a fraction of the beam the fixed-frequency coupling field can dress (Section S6 of the Supporting Information).

The numbers above are predictions of a spectroscopically calibrated model, and direct detection of the prepared population is the next experimental step. 
Two points complete the picture. First, the predictions refer to the molecules that the fixed-frequency coupling field dresses. At the transverse collimation of the present beam only about a fifth of the probed molecules fall inside that class (Section S2 of the Supporting Information), so tightening the transverse velocity distribution is what converts a high transfer probability into a high useful yield. Second, whether any of this matters depends on whether the conditions it needs can be met in a chemical experiment. They are easy to state: the molecule must have a ladder of three rovibrational levels joined by infrared-active transitions, the coupling Rabi frequency must exceed the excitation linewidth of the dressed branches, a few megahertz here, and continuous-wave light of a few tens of milliwatts must be available at the right wavelength. Nothing in the argument requires a permanent dipole moment, which is why a nonpolar species such as CO$_2$ can be used as the demonstration case, and the mid-infrared optical parametric oscillators employed here cover a wavelength range wide enough to address many molecular vibrations. Once prepared, the ensemble is free of external fields, the relevant lifetime is that of a vibrational level in the electronic ground state rather than of an electronically excited state, and at the beam velocity used here the $2.7$~ms lifetime~\cite{GORDON2026HITRAN2024} corresponds to meters of flight, ample for transporting the oriented molecules into a scattering or collision region. Together with universality across polar and nonpolar species, these properties meet the four conditions identified at the outset: magnetic-sublevel selection, complete population transfer, a long-lived prepared state, and preparation into a field-free environment.

In summary, detuning-controlled dressing makes the composition of a dressed rovibrational state a quantity one can reach experimentally, and the resulting adiabatic passage lifts the 50\% ceiling that resonant dressing imposes. The $^{13}$CO$_2$ measurements reported here establish the spectroscopic basis for this control: they resolve the dressed doublet of one polarization-selected magnetic sublevel and show that a single branch can be excited with 97\% selectivity. With the mapping stage added by the detuned field, a calibrated model then predicts near-complete transfer into one $M$ sublevel of a vibrational state that lives for milliseconds. Millisecond-lived oriented molecules of a nonpolar species, prepared in flight and free of external fields, would bring to nonpolar systems the incident-orientation control that has so far been limited to polar molecules; existing orientation-resolved work on nonpolar molecules, such as CO$_2$ scattering from liquid surfaces~\cite{PerkinsNesbitt2008, Perkins2010JPCA}, has characterized the orientation of the scattered products rather than preparing that of the incoming ensemble.

\begin{acknowledgement}
The authors thank T.~Xia and C.-L.~Zou for useful discussions. This work was jointly supported by the Chinese Academy of Sciences (Grant No.~YSBR-055), the National Natural Science Foundation of China (Grant Nos.~22241302, 12393825, 12504569), the China Postdoctoral Science Foundation (Grant No.~2024M753081), and the Innovation Program for Quantum Science and Technology (Grant No.~2021ZD0303102).
The authors used a generative language model to assist with language editing during the preparation of this manuscript; all scientific content, analysis and conclusions are the authors' own, and they reviewed and approved the final text.
\end{acknowledgement}

\noindent\textbf{Supporting Information Available:} the optical layout and frequency stabilization of the two mid-infrared optical parametric oscillators (S1); the data-processing and spectral-fitting procedure, including extraction of the dressed doublet and of the dressed velocity fraction (S2); the three-level Lindblad model, velocity averaging and the Clebsch--Gordan factors used in the simulations (S3); the adiabaticity condition and the rotation of the dressed basis (S4); the five-factor decomposition of the transfer probability (S5); and the dependence of the transfer on coupling power, detuning and velocity spread, with robustness maps (S6). This material is available free of charge via the Internet at http://pubs.acs.org.

\section*{Author Contributions}
M.-Y. Yu: data curation, simulation, visualization, and writing (original draft and editing); Y.-N. Lv: theoretical model development and writing (original draft and editing); C.-F. Cheng: project administration and writing (original draft and editing); S.-M. Hu: supervision and editing.

\section*{Notes}
The authors declare no competing financial interests.

\paragraph*{Data Availability Statement.} The experimental spectra and the parameters used in the simulations are available from the corresponding authors upon reasonable request. The fitting and simulation parameters are tabulated in the Supporting Information (Table~S2), and the raw spectra together with the simulation scripts will be deposited in a public repository upon acceptance.

\bibliography{AT}

@PREAMBLE{
 "\providecommand{\noopsort}[1]{}" 
 # "\providecommand{\singleletter}[1]{#1}%" 
}

@article{zhangOE2020,
  author    = {Zhang, Z.-T. and Cheng, C.-F. and Sun, Y. R. and Liu, A.-W. and Hu, S.-M.},
  title     = {Cavity ring-down spectroscopy based on a comb-locked optical parametric oscillator source},
  journal   = {Opt. Express},
  volume    = {28},
  number    = {19},
  pages     = {27600--27607},
  year      = {2020},
  publisher = {Optica Publishing Group},
  doi       = {10.1364/OE.404944}
}

@article{Tan2024CJCP-OPO,
  author    = {Tan, Yan-Dong and Chen, Jinjin and Zhou, Yong and Cheng, Cun-Feng and Hu, Shui-Ming},
  title     = {Frequency-stabilized mid-infrared laser source for precision spectroscopy},
  journal   = {Chin. J. Chem. Phys.},
  volume    = {37},
  number    = {2},
  pages     = {147--152},
  month     = apr,
  year      = {2024},
  issn      = {2327-2244},
  publisher = {American Institute of Physics},
  doi       = {10.1063/1674-0068/cjcp2312137}
}

@article{lemeshko2013,
  author  = {Lemeshko, M. and Krems, R. V. and Doyle, J. M. and Kais, S.},
  title   = {Manipulation of molecules with electromagnetic fields},
  journal = {Mol. Phys.},
  volume  = {111},
  pages   = {1648--1682},
  year    = {2013},
  doi     = {10.1080/00268976.2013.813595}
}

@article{koch2019,
  author  = {Koch, C. P. and Lemeshko, M. and Sugny, D.},
  title   = {Quantum control of molecular rotation},
  journal = {Rev. Mod. Phys.},
  volume  = {91},
  pages   = {035005},
  year    = {2019},
  doi     = {10.1103/RevModPhys.91.035005}
}

@article{brooks1976,
  author  = {Brooks, P. R.},
  title   = {Reactions of oriented molecules},
  journal = {Science},
  volume  = {193},
  pages   = {11--16},
  year    = {1976},
  doi     = {10.1126/science.193.4247.11}
}

@article{polanyi1987,
  author  = {Polanyi, J. C.},
  title   = {Some concepts in reaction dynamics},
  journal = {Angew. Chem. Int. Ed.},
  volume  = {26},
  pages   = {952--971},
  year    = {1987},
  doi     = {10.1002/anie.198709521}
}

@article{vallance2011,
  author  = {Vallance, C.},
  title   = {Generation, characterisation, and applications of atomic and molecular alignment and orientation},
  journal = {Phys. Chem. Chem. Phys.},
  volume  = {13},
  pages   = {14427--14441},
  year    = {2011},
  doi     = {10.1039/C1CP21037H}
}

@article{yoder2010,
  author  = {Yoder, B. L. and Bisson, R. and Beck, R. D.},
  title   = {Steric effects in the chemisorption of vibrationally excited methane on {Ni}(100)},
  journal = {Science},
  volume  = {329},
  pages   = {553--556},
  year    = {2010},
  doi     = {10.1126/science.1191751}
}

@article{hou1997,
  author  = {Hou, H. and Gulding, S. J. and Rettner, C. T. and Wodtke, A. M. and Auerbach, D. J.},
  title   = {The Stereodynamics of a Gas-Surface Reaction},
  journal = {Science},
  volume  = {277},
  number  = {5322},
  pages   = {80--82},
  year    = {1997},
  doi     = {10.1126/science.277.5322.80},
  url     = {https://doi.org/10.1126/science.277.5322.80}
}

@article{kramer1965,
  author  = {Kramer, K. H. and Bernstein, R. B.},
  title   = {Focusing and orientation of symmetric-top molecules with the electric six-pole field},
  journal = {J. Chem. Phys.},
  volume  = {42},
  pages   = {767--770},
  year    = {1965},
  doi     = {10.1063/1.1696004}
}

@article{cho1991,
  author  = {Cho, V. A. and Bernstein, R. B.},
  title   = {Tight focusing of beams of polar polyatomic molecules via the electrostatic hexapole lens},
  journal = {J. Phys. Chem.},
  volume  = {95},
  pages   = {8129--8136},
  year    = {1991},
  doi     = {10.1021/j100174a022}
}

@article{loison1995,
  author  = {Loison, J. C. and Durand, A. and Bazalgette, G. and White, R. and Audouard, E. and Vigu\'e, J.},
  title   = {Molecular axis orientation by the brute-force method},
  journal = {J. Phys. Chem.},
  volume  = {99},
  pages   = {13591--13596},
  year    = {1995},
  doi     = {10.1021/j100037a003}
}

@article{pullman1990,
  author  = {Pullman, D. P. and Friedrich, B. and Herschbach, D. R.},
  title   = {Facile alignment of molecular rotation in supersonic beams},
  journal = {J. Chem. Phys.},
  volume  = {93},
  pages   = {3224--3236},
  year    = {1990},
  doi     = {10.1063/1.458855}
}

@article{aquilanti1994,
  author  = {Aquilanti, V. and Ascenzi, D. and Cappelletti, D. and Pirani, F.},
  title   = {Velocity dependence of collisional alignment of oxygen molecules in gaseous expansions},
  journal = {Nature},
  volume  = {371},
  pages   = {399--402},
  year    = {1994},
  doi     = {10.1038/371399a0}
}

@article{pirani2001,
  author  = {Pirani, F. and Cappelletti, D. and Bartolomei, M. and Aquilanti, V. and Scotoni, M. and Vescovi, M. and Ascenzi, D. and Bassi, D.},
  title   = {Orientation of benzene in supersonic expansions, probed by {IR}-laser absorption and by molecular beam scattering},
  journal = {Phys. Rev. Lett.},
  volume  = {86},
  pages   = {5035--5038},
  year    = {2001},
  doi     = {10.1103/PhysRevLett.86.5035}
}

@article{kim1996,
  author  = {Kim, W. and Felker, P. M.},
  title   = {Spectroscopy of pendular states in optical-field-aligned species},
  journal = {J. Chem. Phys.},
  volume  = {104},
  pages   = {1147--1150},
  year    = {1996},
  doi     = {10.1063/1.470770}
}

@article{sakai1999,
  author  = {Sakai, H. and Safvan, C. P. and Larsen, J. J. and Hilligsoe, K. M. and Hald, K. and Stapelfeldt, H.},
  title   = {Controlling the alignment of neutral molecules by a strong laser field},
  journal = {J. Chem. Phys.},
  volume  = {110},
  pages   = {10235--10238},
  year    = {1999},
  doi     = {10.1063/1.478039}
}

@article{larsen1999,
  author  = {Larsen, J. J. and Sakai, H. and Safvan, C. P. and Wendt-Larsen, I. and Stapelfeldt, H.},
  title   = {Aligning molecules with intense nonresonant laser fields},
  journal = {J. Chem. Phys.},
  volume  = {111},
  pages   = {7774--7781},
  year    = {1999},
  doi     = {10.1063/1.480112}
}

@article{stapelfeldt2003,
  author  = {Stapelfeldt, H. and Seideman, T.},
  title   = {Colloquium: Aligning molecules with strong laser pulses},
  journal = {Rev. Mod. Phys.},
  volume  = {75},
  pages   = {543--557},
  year    = {2003},
  doi     = {10.1103/RevModPhys.75.543}
}

@book{scoles1988,
  author    = {Scoles, G.},
  title     = {Atomic and Molecular Beam Methods},
  publisher = {Oxford University Press},
  address   = {New York},
  year      = {1988}
}

@article{spano2001,
  author  = {Spano, F. C.},
  title   = {Theory of sub-{D}oppler {A}utler-{T}ownes splitting in molecules: Alignment and orientation of the angular momentum in nonpolar molecules},
  journal = {J. Chem. Phys.},
  volume  = {114},
  pages   = {276--288},
  year    = {2001},
  doi     = {10.1063/1.1328381}
}

@article{stahovich2024,
  author  = {Stahovich, J. T. and Rowe, B. A. and Huennekens, J. and Lyyra, A. M. and Ahmed, E. H.},
  title   = {Molecular angular momentum orientation using dressed states created by laser radiation},
  journal = {Phys. Rev. Lett.},
  volume  = {133},
  pages   = {263601},
  year    = {2024},
  doi     = {10.1103/PhysRevLett.133.263601}
}

@article{Carr2009NJP,
  author    = {Carr, Lincoln D. and DeMille, David and Krems, Roman V. and Ye, Jun},
  title     = {Cold and ultracold molecules: science, technology and applications},
  journal   = {New J. Phys.},
  volume    = {11},
  number    = {5},
  pages     = {055049},
  year      = {2009},
  month     = {may},
  publisher = {IOP Publishing},
  doi       = {10.1088/1367-2630/11/5/055049},
  url       = {https://doi.org/10.1088/1367-2630/11/5/055049}
}

@article{Krems2008PCCP,
  author  = {Krems, R. V.},
  title   = {Cold controlled chemistry},
  journal = {Phys. Chem. Chem. Phys.},
  volume  = {10},
  number  = {28},
  pages   = {4079--4092},
  year    = {2008},
  doi     = {10.1039/b802322k},
  url     = {https://doi.org/10.1039/b802322k}
}

@article{Ni2008Science,
  author    = {Ni, K.-K. and Ospelkaus, S. and de Miranda, M. H. G. and Pe'er, A. and Neyenhuis, B. and Zirbel, J. J. and Kotochigova, S. and Julienne, P. S. and Jin, D. S. and Ye, J.},
  title     = {A High Phase-Space-Density Gas of Polar Molecules},
  journal   = {Science},
  volume    = {322},
  number    = {5899},
  pages     = {231--235},
  year      = {2008},
  doi       = {10.1126/science.1163861},
  url       = {https://doi.org/10.1126/science.1163861},
  eprint    = {https://www.science.org/doi/pdf/10.1126/science.1163861}
}

@article{DeMille2002PRL,
  title     = {Quantum Computation with Trapped Polar Molecules},
  author    = {DeMille, D.},
  journal   = {Phys. Rev. Lett.},
  volume    = {88},
  number    = {6},
  pages     = {067901},
  numpages  = {4},
  year      = {2002},
  month     = {jan},
  publisher = {American Physical Society},
  doi       = {10.1103/PhysRevLett.88.067901},
  url       = {https://doi.org/10.1103/PhysRevLett.88.067901}
}

@article{Friedrich1991Nature,
  author  = {Friedrich, Bretislav and Herschbach, Dudley R.},
  title   = {Spatial orientation of molecules in strong electric fields and evidence for pendular states},
  journal = {Nature},
  volume  = {353},
  pages   = {412--414},
  year    = {1991},
  doi     = {10.1038/353412a0}
}

@article{Goban2008PRL,
  title     = {Laser-Field-Free Molecular Orientation},
  author    = {Goban, Akihisa and Minemoto, Shinichirou and Sakai, Hirofumi},
  journal   = {Phys. Rev. Lett.},
  volume    = {101},
  number    = {1},
  pages     = {013001},
  numpages  = {4},
  year      = {2008},
  month     = {jun},
  publisher = {American Physical Society},
  doi       = {10.1103/PhysRevLett.101.013001},
  url       = {https://doi.org/10.1103/PhysRevLett.101.013001}
}

@article{Ghafur2009NP,
  author    = {Ghafur, Omair and Rouz\'ee, Arnaud and Gijsbertsen, Arjan and Siu, Wing Kiu and Stolte, Steven and Vrakking, Marc J. J.},
  title     = {Impulsive orientation and alignment of quantum-state-selected {NO} molecules},
  journal   = {Nat. Phys.},
  volume    = {5},
  pages     = {289--293},
  year      = {2009},
  publisher = {Nature Publishing Group},
  doi       = {10.1038/nphys1225}
}

@article{Vainio2016PCCP,
  author    = {Vainio, M. and Halonen, L.},
  title     = {Mid-infrared optical parametric oscillators and frequency combs for molecular spectroscopy},
  journal   = {Phys. Chem. Chem. Phys.},
  volume    = {18},
  number    = {6},
  pages     = {4266--4294},
  year      = {2016},
  month     = {feb},
  issn      = {1463-9076},
  doi       = {10.1039/c5cp07052j},
  url       = {https://doi.org/10.1039/c5cp07052j}
}

@article{GORDON2026HITRAN2024,
  author  = {Gordon, I. E. and Rothman, L. S. and others},
  title   = {The {HITRAN2024} molecular spectroscopic database},
  journal = {J. Quant. Spectrosc. Radiat. Transf.},
  volume  = {353},
  pages   = {109807},
  year    = {2026},
  issn    = {0022-4073},
  doi     = {10.1016/j.jqsrt.2026.109807}
}

@book{Herzberg1945,
  author    = {Herzberg, G.},
  title     = {Molecular Spectra and Molecular Structure, Vol. 2},
  publisher = {Springer},
  address   = {New York},
  year      = {1945}
}

@article{CohenTannoudjiHaroche1969,
  author  = {Cohen-Tannoudji, C. and Haroche, S.},
  title   = {Absorption et diffusion de photons optiques par un atome
             en interaction avec des photons de radiofr\'equence},
  journal = {J. Phys. France},
  volume  = {30},
  pages   = {153--168},
  year    = {1969},
  doi     = {10.1051/jphys:01969003002-3015300}
}

@article{Mollow1969,
  author  = {Mollow, B. R.},
  title   = {Power Spectrum of Light Scattered by Two-Level Systems},
  journal = {Phys. Rev.},
  volume  = {188},
  pages   = {1969--1975},
  year    = {1969},
  doi     = {10.1103/PhysRev.188.1969}
}

@article{Bergmann1998,
  author  = {Bergmann, K. and Theuer, H. and Shore, B. W.},
  title   = {Coherent Population Transfer among Quantum States of
             Atoms and Molecules},
  journal = {Rev. Mod. Phys.},
  volume  = {70},
  pages   = {1003--1025},
  year    = {1998},
  doi     = {10.1103/RevModPhys.70.1003}
}

@article{Perkins2010JPCA,
  author  = {Perkins, B. G., Jr. and Nesbitt, D. J.},
  title   = {Stereodynamics at the Gas--Liquid Interface: Orientation and Alignment of {CO}$_{2}$ Scattered from Perfluorinated Liquid Surfaces},
  journal = {J. Phys. Chem. A},
  volume  = {114},
  number  = {3},
  pages   = {1398--1410},
  year    = {2010},
  doi     = {10.1021/jp907022u}
}

@article{PerkinsNesbitt2008,
  author  = {Perkins, B. G., Jr. and Nesbitt, D. J.},
  title   = {Stereodynamics in State-Resolved Scattering at the Gas--Liquid Interface},
  journal = {Proc. Natl. Acad. Sci. U.S.A.},
  volume  = {105},
  number  = {35},
  pages   = {12684--12689},
  year    = {2008},
  doi     = {10.1073/pnas.0800401105}
}

\end{document}


\maketitle

\renewcommand\thefigure{S\arabic{figure}}\setcounter{figure}{0}
\renewcommand\thetable{S\arabic{table}}\setcounter{table}{0}
\renewcommand\theequation{S\arabic{equation}}\setcounter{equation}{0}

\section*{S1. Laser frequency control}

\begin{figure}[H]
    \centering
    \includegraphics[width=0.8\textwidth]{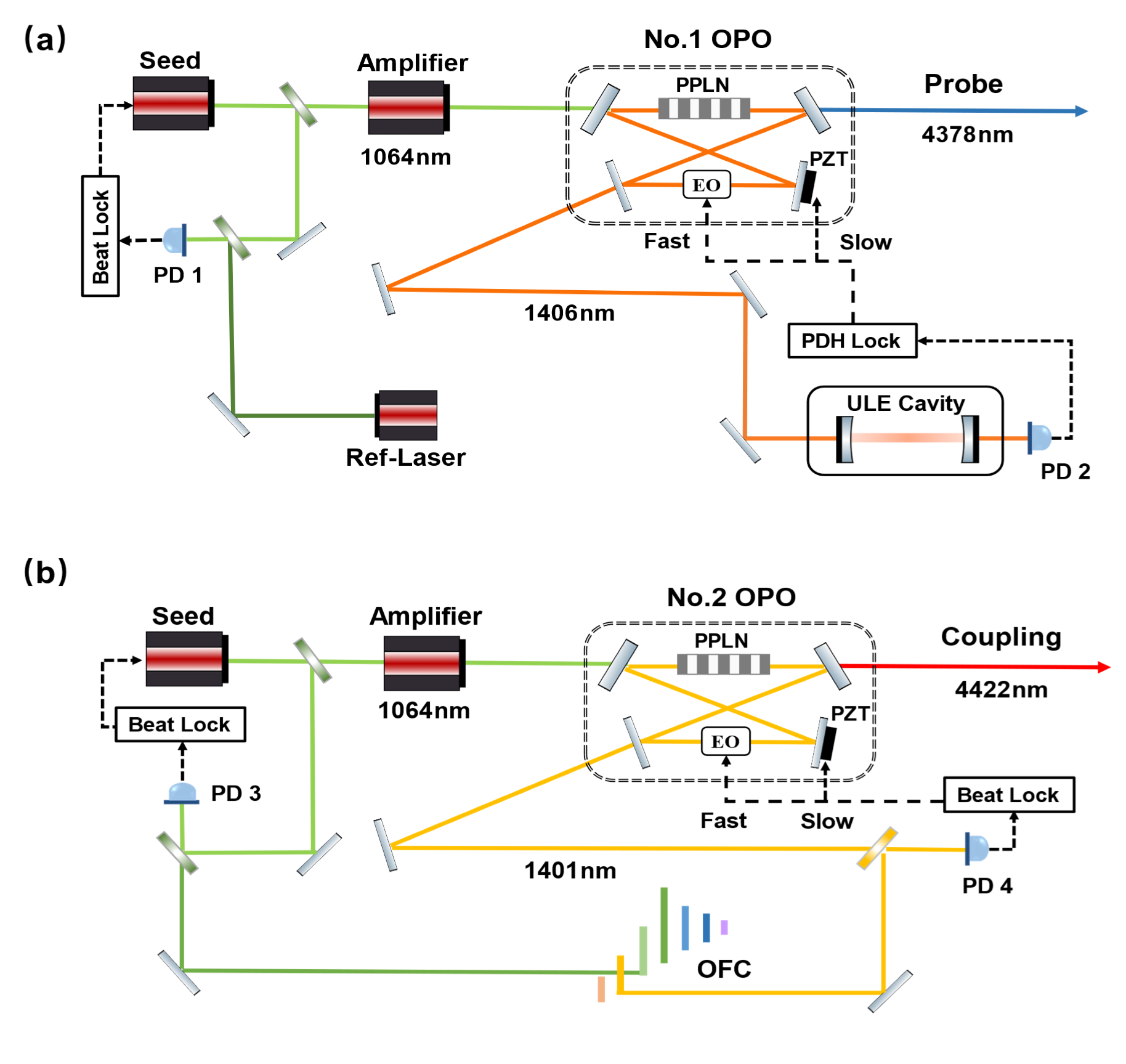}
    \caption{Schematic of the optical setup. Upper and lower panels show the frequency-locking chains of the probe OPO (ULE cavity plus PDH) and of the coupling OPO (optical frequency comb).}
    \label{Fig:SM_setup}
\end{figure}

The molecular beam intersects two counter-propagating infrared laser beams at the interaction region: the probe and coupling fields. Both are derived from continuous-wave optical parametric oscillators (OPOs), as shown in Figure~\ref{Fig:SM_setup}. The coupling laser (No.~2 OPO), resonant with the vibrational transition near 4422~nm, delivers up to 200~mW at the OPO outlet and 60~mW at the interaction region. The probe laser (No.~1 OPO), interrogating the band near 4378~nm, is held below 10~$\mu$W so that detection does not perturb the populations prepared by the coupling field. Each beam passes through an independent quarter-wave plate, permitting the polarization state (left- or right-circular) to be set individually, and is focused to a diameter of 1~mm. The transmitted probe beam is isolated by a polarizing beam splitter (PBS) and detected with a HgCdTe photodiode (PD).

A key requirement of the experiment is that the coupling field delivers sufficient power density (per unit linewidth) to induce splittings that exceed the probe spectral resolution. At the same time the probe field must remain weak enough to avoid redistributing populations or inducing dark-state resonances, while being spectrally narrow enough to resolve the resulting structure. Meeting these conditions demands sub-MHz frequency control of the mid-infrared OPO sources. Each OPO is stabilized by locking its pump and signal fields simultaneously, the absolute frequency of the mid-infrared idler output being inherited from these locks. For the probe OPO the signal light was pre-stabilized to an ultra-low-expansion (ULE) cavity by the Pound--Drever--Hall technique (Figure~\ref{Fig:SM_setup}, upper panel); the pump beam was then phase-locked to a reference laser also stabilized to a ULE cavity, with the linewidth compressed to about 10~kHz. This configuration permits wide tuning of the pump frequency and hence continuous scanning of the probe idler across the resonance. For the coupling OPO, pump and signal beams are locked directly to an optical frequency comb (Figure~\ref{Fig:SM_setup}, lower panel), giving absolute frequency calibration at a selected beat-note offset. Further details of the OPO architecture and locking electronics are given in previous work.

\section*{S2. Data processing and spectral fitting}

This section details the processing of the absorption spectra and the extraction of the Autler--Townes (AT) splitting. We use the spectrum recorded at a coupling Rabi frequency $\Omega_{c,M^\prime}/2\pi = 2.94$~MHz as a representative example, i.e. the data set at this Rabi frequency in Figure~2b of the main text; the identical procedure was applied to every spectrum shown in Figures~2 and 3 of the main text.

The absorption spectrum is decomposed into a Gaussian Doppler background, accounting for the velocity-averaged absorption of molecules not dressed by the coupling field, and an AT doublet arising from the dressed sub-ensemble:
\begin{equation}
\begin{aligned}
A(\nu) &= A_{\mathrm{G}}(\nu) + A_{\mathrm{AT}}(\nu),\\
A_{\mathrm{G}}(\nu) &= A_{\mathrm{g}}\,\exp\!\left[-\frac{(\nu-\nu_c)^2}{2\sigma_{\mathrm{g}}^2}\right] + b,\\
A_{\mathrm{AT}}(\nu) &= \frac{A_{\mathrm{at}}\,(1+a)}{1+4(\nu-\nu_c-\Delta\nu/2)^2/\gamma^2} +
\frac{A_{\mathrm{at}}\,(1-a)}{1+4(\nu-\nu_c+\Delta\nu/2)^2/\gamma^2}.
\end{aligned}
\end{equation}
Here $\Delta\nu$ is the doublet separation, $\gamma$ the common Lorentzian full width at half maximum of the two components, and $a$ their amplitude asymmetry. The Doppler width was fixed at $\sigma_g=2.875$~MHz (FWHM $=6.77$~MHz), matching the fit of the coupling-free spectrum in Figure~2a of the main text. Free parameters were obtained by constrained nonlinear least-squares minimization of $A(\nu)$ to the $N$ data points.

\begin{figure}[H]
    \centering
    \includegraphics[width=0.8\textwidth]{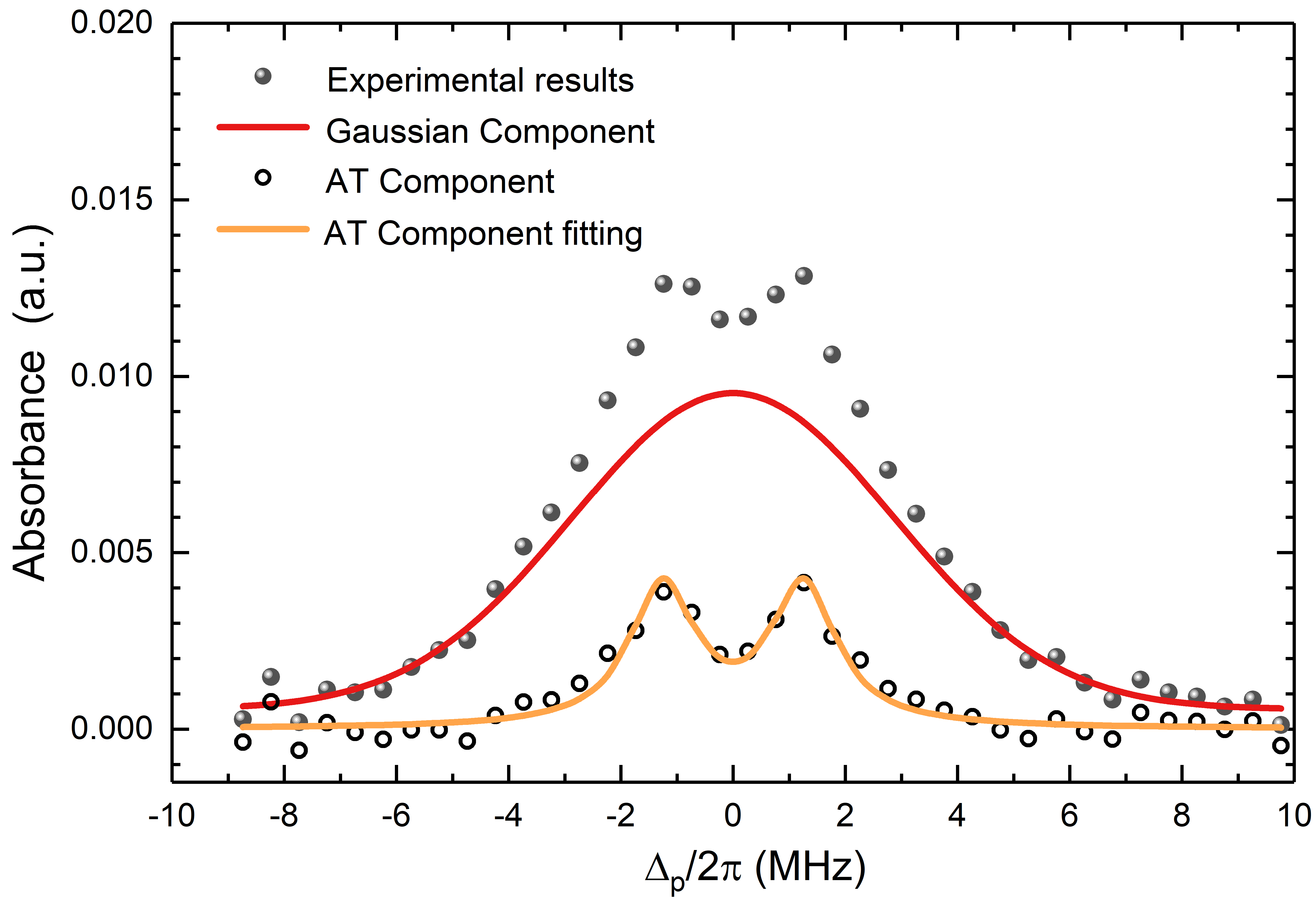}
    \caption{Decomposition of a measured absorption spectrum of the $^{13}$CO$_2$ R(0) rovibrational transition (orange trace of Figure~2b of the main text). Gray filled circles: measured absorption $A(\nu)$. Red curve: fitted Gaussian Doppler background $A_{\mathrm{G}}(\nu)$. Open circles: extracted AT component $A(\nu)-A_{\mathrm{G}}(\nu)$. Orange curve: fitted AT doublet $A_{\mathrm{AT}}(\nu)$. The fitted separation is $\Delta\nu=R_{M}=2.51\pm0.15$~MHz (ordinary-frequency units, $R_M$ itself is an angular frequency, 1$\sigma$ standard error), equal to $\Omega_{c,M^\prime}/2\pi$ at resonance.}
    \label{Fig:fitting}
\end{figure}
Reported uncertainties are the one-standard-deviation standard errors obtained from the fit covariance matrix. For the spectrum in Figure~\ref{Fig:fitting} the fit returns $R^2=0.9973$ (adjusted $R^2=0.9967$), a separation $\Delta\nu=R_M=2.51\pm0.15$~MHz and a negligible asymmetry $a=-0.05$, as expected for resonant coupling ($\Delta_c=0$). Two quoted coupling strengths should be distinguished. The value $\Omega_{c,M^\prime}/2\pi=2.94$~MHz at the start of this section is what the measured beam waist and power give; the fit of the same spectrum returns $R_M/2\pi=2.51\pm0.15$~MHz. The roughly 15\% difference measures the accuracy of the absolute power-density calibration. All simulations below are anchored to the fitted spectra rather than to the nominal value, so the difference enters nowhere else.

As a consistency check, the integrated absorption of the full measured spectrum with the coupling field on agrees with that of the coupling-free spectrum to within 2\% when both are integrated over the same $\pm9.5$~MHz window. This is expected in the weak-probe limit, where the coupling field changes neither the number of molecules probed nor the total absorption cross-section, and it also validates the data-processing and fitting procedures. At the coupling strength of Figure~\ref{Fig:fitting} the dressed velocity class accounts for roughly 22\% of the total integrated absorption, indicating that a substantial fraction of the beam is engaged by the coupling field and thereby addressable for state-selective manipulation.

\section*{S3. Theoretical model and simulation methods}

A ladder-type three-level system corresponding to the experiment is studied, with the three states denoted $\{|0\rangle,|1\rangle,|2\rangle\}$, where $\lvert0\rangle=\lvert0,0\rangle$, $\lvert1\rangle=\lvert1,-1\rangle$ and $\lvert2\rangle=\lvert2,0\rangle$. The weak probe field couples $|0\rangle\leftrightarrow|1\rangle$ and the coupling field drives $|1\rangle\leftrightarrow|2\rangle$. The molecular velocity is written $\mathbf{v}=(v_z,v_y)$, with $v_z$ along the laser propagation direction giving the Doppler shifts and $v_y$ setting the transit time through the Gaussian beams.

In the rotating-wave approximation and in the interaction picture, the Hamiltonian for a molecule with velocity $(v_y,v_z)$ is
\begin{align}
H(t;v_y,v_z) = {} & \Delta_p(v_z)|1\rangle\langle1| + \left[\Delta_p(v_z)+\Delta_c(v_z)\right] |2\rangle\langle2| \nonumber\\
& + \Big(\frac{\Omega_p(t;v_y)}{2} |1\rangle\langle0| + \frac{\Omega_c(t;v_y)}{2}|2\rangle\langle1|+\mathrm{H.c.}\Big),
\end{align}
where $\Delta_p(v_z)=\omega_{10}-\omega_p-k_p v_z$ and $\Delta_c(v_z)=\omega_{21}-\omega_{c}+k_c v_z$ for the counter-propagating configuration used here. Population dynamics are obtained from the Lindblad master equation
\begin{equation}
\dot{\rho} = -i[H,\rho] + \sum_\mu \left(
L_\mu\rho L_\mu^\dagger -\frac{1}{2} \left\{L_\mu^\dagger L_\mu,\rho \right\} \right),
\end{equation}
with collapse operators $L_{10}=\sqrt{A_{10}}\,|0\rangle\langle1|$ and $L_{21}=\sqrt{A_{21}}\,|1\rangle\langle2|$ accounting for spontaneous emission.

Selection rules and channel assignment. With the polarizations of Figure~1(a) of the main text, the $\sigma^-$ probe drives only $\Delta M=-1$ transitions, so from the unique $M=0$ ground state it populates only $\ket{1,-1}$; the $\sigma^+$ coupling then addresses only the $M=-1\to M'=0$ channel. The measured doublet is therefore assigned to $(M,M')=(-1,0)$. The other channels are doubly excluded: they are not driven by the polarization combination used, and, through the $M$-dependence of $R_M$ arising from the coefficients of Table~\ref{tab:app_CG}, they would in any case appear at different probe frequencies, as stated in the main text.

For a molecule traversing a Gaussian spatial envelope, the Rabi frequencies vary along the trajectory according to
\begin{equation}
\Omega_j(t;v_y) = \Omega_j^{\max} C_{J M_J,1q}^{J'M_J'} \exp\left[ -\frac{(t-t_c)^2} {2\sigma_{t}^2(v_y)}\right],
\end{equation}
with $\sigma_t(v_y)=\sigma_r/v_y$, $\sigma_r$ the spatial width of the beam, and $t_c$ the center of the pulse. Here $\Omega_j^{\max} = - d\cdot E_{\max}$, $E_{\max}$ being the peak field amplitude and $d$ the transition dipole moment, while $C_{J M_J,1q}^{J'M_J'} = \langle J M_J;1q|J'M_J'\rangle$ is the Clebsch--Gordan coefficient of the transition $\lvert J,M\rangle\to\lvert J', M'\rangle$ and $q=0,\pm1$ labels the polarization component. Allowed sublevels obey $M_J'=M_J+q$, with $q=0$ for a $\pi$ component and $q=\pm1$ for the two circular components. Clebsch--Gordan coefficients for the coupling transition $|J=1\rangle\to|J'=2\rangle$ are listed in Table~\ref{tab:app_CG}; for each final sublevel $M_J'$ the squared coefficients summed over all allowed $(M_J,q)$ pairs satisfy $\sum_{M_J,q}|C_{1 M_J,1q}^{2 M_J'}|^2=1$, consistent with their normalization.
\begin{table}[htbp]
\centering
\caption{\label{tab:app_CG}
Clebsch--Gordan coefficients $C_{1 M_J,1q}^{2 M_J'}=\langle 1\,M_J;1\,q|2\,M_J'\rangle$ for the $|J=1\rangle\to|J'=2\rangle$ coupling transition, with squared moduli $|C|^2$ given as relative transition strengths.
}
\begin{tabular}{cccccccccc}
\hline
$M_J$ & $-1$ & $0$ & $+1$ & $-1$ & $0$ & $+1$ & $-1$ & $0$ & $+1$ \\
$q$ & $-1$ & $-1$ & $-1$ & $0$ & $0$ & $0$ & $+1$ & $+1$ & $+1$ \\
$M_J'=M_J+q$ & $-2$ & $-1$ & $0$ & $-1$ & $0$ & $+1$ & $0$ & $+1$ & $+2$ \\
polarization & $\sigma^-$ & $\sigma^-$ & $\sigma^-$ & $\pi$ & $\pi$ & $\pi$ & $\sigma^+$ & $\sigma^+$ & $\sigma^+$ \\
$C_{1 M_J,1q}^{2 M_J'}$ & $1$ & $1/\sqrt{2}$ & $1/\sqrt{6}$ & $1/\sqrt{2}$ & $\sqrt{2/3}$ & $1/\sqrt{2}$ & $1/\sqrt{6}$ & $1/\sqrt{2}$ & $1$ \\
$|C|^2$ & $1$ & $1/2$ & $1/6$ & $1/2$ & $2/3$ & $1/2$ & $1/6$ & $1/2$ & $1$ \\
\hline
\end{tabular}
\end{table}

\begin{table}[htbp]
\centering
\caption{\label{tab:params}
Provenance of the model parameters: each entry states whether the value comes from an independent measurement, from the literature, or from the spectral fit of Section~S2. The simulations of Figures~2c and~4 of the main text use the fitted spectral parameters together with independently measured beam and velocity data.
}
\small
\begin{tabular}{@{}llp{0.44\textwidth}@{}}
\hline
Parameter & Value & Provenance \\
\hline
$A_{10}$, $A_{21}$ (spontaneous rates) & $1/(2.7\,\mathrm{ms})$ & HITRAN database, as cited in the main text \\
$\sigma_r$ (Gaussian beam radius) & $0.5$~mm & measured 1~mm beam diameter \\
$\bar v_y$ & $700$~m/s & time-of-flight measurement \\
$3\sigma_{v_y}$ & $40$~m/s & beam diagnostics \\
$\sigma_{v_z}$ & $0.7$~m/s & dressed sub-ensemble width (S6) \\
$\Omega_{c,M'}(P_c)$ & calibrated & power meter and waist, cross-checked with the fit (S2) \\
$\Gamma_{\mathrm{leak}}$ & $0$ & negligible on $\mu$s timescales \\
$C^{J'M_J'}_{JM_J,1q}$ & Table~S1 & angular-momentum algebra \\
\hline
\end{tabular}
\end{table}


Both velocity components are taken as Gaussian distributed with widths $\sigma_{v_y}$ and $\sigma_{v_z}$,
\begin{equation}
f_a(v_a) = \frac{1} {\sqrt{2\pi}\sigma_{v_a}}\exp\left[-\frac{(v_a-\bar{v}_a)^2}{2\sigma_{v_a}^2}\right],\quad a\in (y,z),
\end{equation}
and the ensemble-averaged population of state $\lvert a\rangle$ is
\begin{equation}
\langle \rho_{aa}\rangle = \int dv_y\,dv_z\, f_y(v_y)f_z(v_z)\rho_{aa}(v_y,v_z).
\end{equation}
The theoretical results shown in Figures~2c and 4 of the main text follow from this model.

\section*{S4. Condition for adiabatic evolution}

Adiabaticity of the $E_{\pm,M}$ evolution is determined by comparing the rate of change of the instantaneous eigenstate, $|\dot{\theta}_{M}(t)|$, with the instantaneous energy gap $R_{M}(t)$. Whenever
\begin{equation}
\eta(t) = \frac{|\dot{\theta}_{M}(t)|}{R_{M}(t)}\ll1,
\label{eq:app_adiabatic_parameter}
\end{equation}
the dressed states evolve adiabatically. With $\theta_M=\arccos\sqrt{(R_M+\Delta_c)/2R_M}$,
\begin{equation}
\dot{\theta}_{M}(t) = \frac{\Delta_c\dot{\Omega}_{c,M'}(t)}{2\left[\Delta_c^2+\Omega_{c,M'}^2(t)\right]},
\end{equation}
and for the Gaussian coupling envelope
\begin{equation}
\dot{\Omega}_{c,M'}(t) = -\frac{t-t_c}{\sigma_t^2}\Omega_{c,M'}(t),
\end{equation}
so that condition~(\ref{eq:app_adiabatic_parameter}) becomes
\begin{align}
\left| \frac{\Delta_{c}}{2R_{M}^{2}(t)}\frac{t-t_{c}}{\sigma_{t}^{2}}\Omega_{c,M'}(t)\right| \ll R_{M}(t).
\label{eq:ad_con}
\end{align}
Adiabatic following is thus guaranteed at the pulse center, where $\dot{\Omega}_{c,M'}(t_c)=0$, in the far wings, where $\Omega_{c,M'}(t)\to0$, and on resonance, $\Delta_c=0$; the dominant nonadiabatic contribution arises on the rising and falling edges of the pulse for $\Delta_c\neq0$.

\begin{figure}[H]
    \centering
    \includegraphics[width=0.8\textwidth]{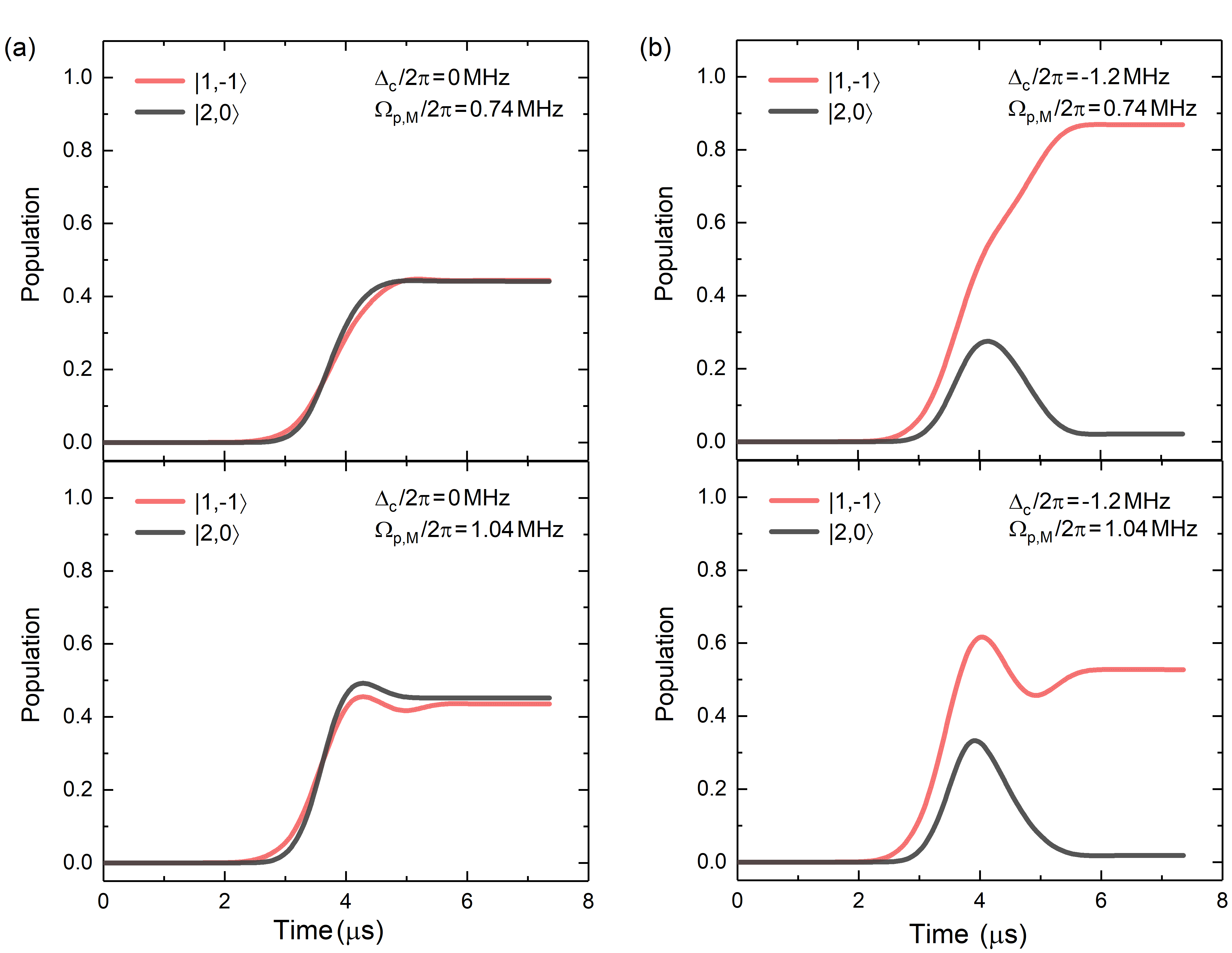}
    \caption{Time evolution of the populations of $|1,-1\rangle$ (red) and $|2,0\rangle$ (black) for various $\Omega_{p,M=-1}$ and $\Delta_c$. (a) Resonant case, $\Delta_c/2\pi=0$~MHz. (b) Coupling detuning $\Delta_c/2\pi=-1.2$~MHz. Upper and lower plots in each panel use peak probe Rabi frequencies of $0.74$ and $1.04$~MHz.}
    \label{FigS:ATDetuning}
\end{figure}

Resonant and detuned cases differ qualitatively. On resonance the mixing angle is constant, $\theta_M(t)=\pi/4$, giving time-independent dressed states $|E_{-,-1}^{(\mathrm{res})}\rangle=(|1,-1\rangle - |2,0\rangle)/\sqrt{2}$ and $|E_{+,-1}^{(\mathrm{res})}\rangle=(|1,-1\rangle+|2,0\rangle)/\sqrt{2}$, with the bare states mixed in a fixed 1:1 ratio throughout the evolution [Figure~\ref{FigS:ATDetuning}(a)]. The mapping factor is therefore capped at $\Pmap=1/2$ and a resonant scheme has no unique bare-state destination. Off resonance the mixing angle varies with $\Omega_{c,M'}(t)$, so the dressed-state basis rotates during the pulse. When $|\dot{\theta}_{M}(t)|\ll R_{M}(t)$, transitions between the branches are suppressed while the bare-state composition of the followed branch still changes. For the negative detuning used in Figure~4 of the main text, the dressed states become pure bare states as $\Omega_{c,M'}(t)\to0$, namely $|E_{+,-1}(t)\rangle\to|1,-1\rangle$ and $|E_{-,-1}(t)\rangle\to|2,0\rangle$, as illustrated in Figure~\ref{FigS:ATDetuning}(b).

For the trajectories actually realized, $\sigma_t=\sigma_r/v_y=0.5$~mm$/700$~m/s $\approx0.71$~$\mu$s and $|\Delta_c|/2\pi=1.2$~MHz give the dimensionless product $\sigma_t|\Delta_c|\approx5.3$. Writing $x=\Omega_{c,M'}(t)/|\Delta_c|$ and $\tau=(t-t_c)/\sigma_t$, condition~(\ref{eq:app_adiabatic_parameter}) reads $\eta(t)=x|\tau|/[2\sigma_t|\Delta_c|(1+x^2)^{3/2}]$, whose maximum over the falling edge ($x\lesssim\sqrt{2}$, $\tau\lesssim2$) is of order $0.05$, comfortably $\ll1$. The minimum energy gap during turnoff is $\min_t R_M(t)=|\Delta_c|$, reached as $\Omega_{c,M'}\to0$; the detuning itself is what keeps the gap open at the end of the pulse.

\section*{S5. Quantifying the population-transfer efficiency}

The quantity of interest is the final population $P_{\mathrm{final}}=\langle 1,-1|\bar{\rho}(t_f)|1,-1\rangle$ obtained from the master equation after velocity averaging. To identify the dominant limitation of each stage, it is decomposed into five diagnostic factors,
\begin{equation}
P_{\mathrm{final}}\simeq\Pload\,\Psel\,\Pad\,\Pmap\,\Psurv.
\end{equation}
The target branch is $E_{+,-1}$ for $\Delta_c<0$, i.e. the branch continuously connected to $\lvert 1,-1\rangle$ as $\Omega_{c,M'=0}\to0$, and its bare-state weights are written
\begin{equation}
\sin^2\theta_M=\frac{R_M-\Delta_c}{2R_M},\qquad
\cos^2\theta_M=\frac{R_M+\Delta_c}{2R_M},
\end{equation}
so that $\sin^2\theta_M\ge 1/2$ for $\Delta_c<0$ and $\sin^2\theta_M\to1$ as $\Omega_{c,M'}\to0$.

$\Pload$ (branch-selective loading). Treating the target branch as an isolated effective two-level system driven with strength $\Omega_{p,\mathrm{eff}}=\Omega_p \sin\theta_{M=-1}$,
\begin{equation}
\Pload\simeq\sin^2(\Theta/2),\qquad
\Theta=\int \Omega_p(t)\,\sin\theta_{M=-1}(t)\,dt .
\end{equation}
It is limited by the probe strength, spectral overlap, and transit time; once $\Theta$ exceeds $\pi$, coherent overdriving reduces $\Pload$ again.

$\Psel$ (branch selection). The target branch $E_{+,M}$ corresponds to the left component of the probe spectrum. At its fitted resonance $\Delta_{p,0}$ the spectral selectivity is $P_{\mathrm{sel}}=A_L(\Delta_{p,0})/[A_L(\Delta_{p,0})+A_R(\Delta_{p,0})]$. For two Lorentzian components with peak heights $h_L$, $h_R$ and separation $R_M$, evaluation at the target peak center gives
\begin{equation}
P_{\mathrm{sel}}^{(L)}\simeq \left[1+\frac{\beta}{1+4(R_M/\Gamma_{\mathrm{other}})^2}\right]^{-1},
\qquad
\beta=\frac{h_R}{h_L},
\end{equation}
where $\Gamma_{\mathrm{other}}$ is the FWHM of the non-target branch expressed in the same angular-frequency units as $R_M$. For $\Delta_c<0$ the target branch carries the larger $|1,M\rangle$ weight and $\beta<1$, so selectivity tends to unity when $R_M\gg\Gamma_{\mathrm{other}}$ or $\beta\to0$. Values reported in Figure~3 of the main text are evaluated from fitted peak heights, separation, and linewidth rather than from the ideal mixing-angle estimate for $\beta$.

For the operating point $\Delta_c/2\pi=-1.2$~MHz the fit returns $P_{\mathrm{sel}}=0.97\pm0.05$. The 1$\sigma$ uncertainty quoted here is the conservative estimate, propagated from the Doppler-background fit and the line-center extraction preceding the doublet fit in S2. The covariance matrix of the direct fit to the Doppler-subtracted doublet alone would give a much smaller formal uncertainty of 2$\times$10$^{-4}$, which is invisible on the scale of Fig.3(c).

$\Pad$ (adiabatic following). Adiabatic following requires condition~(\ref{eq:app_adiabatic_parameter}) to hold throughout the decrease of the coupling field, giving $P_{\mathrm{ad}}\simeq1$. For the following probability we assume ideal loading into the instantaneous dressed state $|E_{+,-1}(t_c)\rangle$ at the pulse maximum, set the probe field to zero, neglect spontaneous emission and dephasing, and propagate the state along the Gaussian falling edge to $t_f=t_c+5\sigma_t$. The probability of remaining on the same branch is
\begin{equation}
p_{\mathrm{ad}} = \left| \left\langle E_{s,-1}(t_f)\right| U(t_f,t_c) \left|E_{s,-1}(t_c)\right\rangle \right|^2 ,
\end{equation}
$U(t_f,t_c)$ being the unitary propagator generated by $H$ with $\Omega_{p,M}=0$. The values $\Pad=\langle p_{\mathrm{ad}}\rangle$ quoted in the main text are averaged over the Gaussian velocity distributions using the parameters of Figure~4 of the main text. The velocity average runs over the dressed sub-ensemble of Section~S6, i.e. the molecules the coupling field actually engages, so the quoted probability characterizes the molecules that enter the transfer rather than the full beam. The propagation here assumes ideal loading and neglects the probe during the falling edge; the full three-level Lindblad evolution including both fields is what Figures~4 of the main text and~\ref{fig:robustness} show.

$\Pmap$ (dressed-to-bare mapping). This is the $\lvert 1,-1\rangle$ weight of the followed branch at the final time,
\begin{equation}
\Pmap = \sin^2\theta_{M=-1}(t_f)=\frac{R_{M=-1}(t_f)-\Delta_c}{2R_{M=-1}(t_f)}
\simeq 1-\frac{\Omega_{c,M'=0}^2(t_f)}{4\Delta_c^2},
\end{equation}
which tends to unity on the trailing edge for $\Delta_c<0$. On resonance $\theta_M=\pi/4$ at all times and $\Pmap=1/2$, which caps single-sublevel transfer in resonant schemes near 50\%.

$\Psurv$ (dissipative survival). Using a no-jump trajectory as a diagnostic,
\begin{equation}
\Psurv\simeq\exp\!\left[-\int\Gamma_{\mathrm{loss}}(t)\,dt\right],
\qquad
\Gamma_{\mathrm{loss}}\simeq A_{10}\sin^2\theta_{M=-1}+A_{21}\cos^2\theta_{M=-1}+\Gamma_{\mathrm{leak}} .
\end{equation}
With millisecond vibrational lifetimes, $\Psurv\simeq1$ on the microsecond timescale of the experiment.

The five-factor product is a diagnostic decomposition of the target pathway rather than an identity for coherent open-system dynamics; the final criterion remains $\langle 1,-1|\bar{\rho}(t_f)|1,-1\rangle$ from the full master equation. For representative parameters at $\Delta_c/2\pi=-1.2$~MHz, $\Omega_{p,M=-1}/2\pi\simeq0.74$~MHz gives an effective area of $\approx1.1\pi$, combining high $\Pload$ with high selection and mapping purity, whereas $\Omega_{p,M=-1}/2\pi\simeq1.04$~MHz gives $\approx1.6\pi$ and overdriving lowers the net transfer.

\section*{S6. Robustness against experimental parameters}

\begin{figure}[H]
    \centering
    \includegraphics[width=0.8\textwidth]{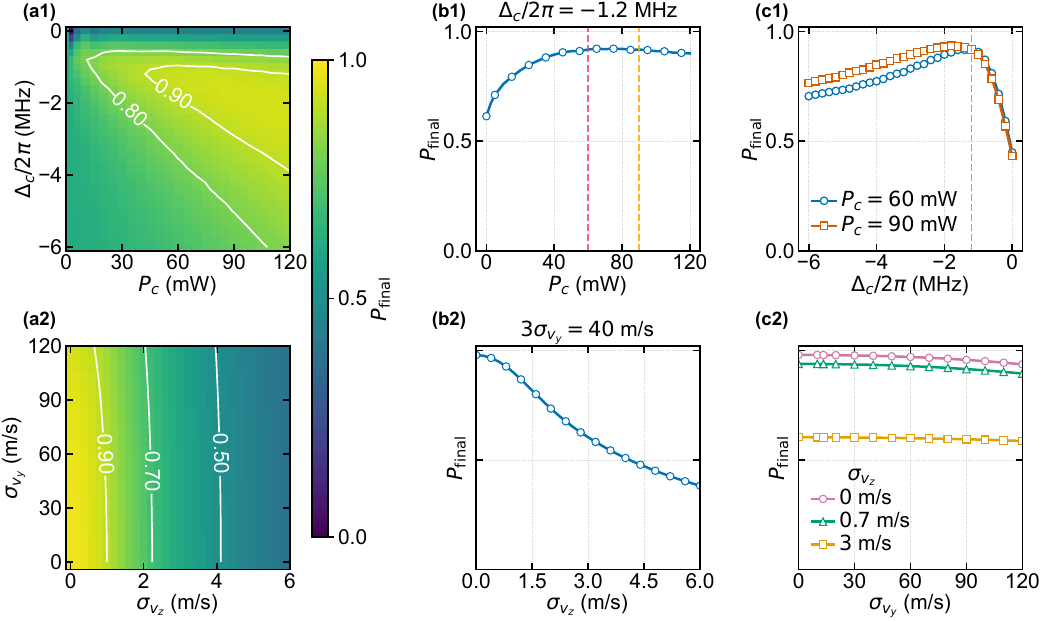}
    \caption{Upper row: two-dimensional control-parameter maps and slices versus coupling power $P_c$ and coupling detuning $\Delta_c$. Lower row: two-dimensional robustness maps and slices versus transverse and longitudinal velocity widths $\sigma_{v_z}$ and $\sigma_{v_y}$. Colors and white contours are obtained from the dissipative three-level master equation with two-dimensional velocity averaging; the color scale is $P_{\mathrm{final}}$. The diamond marks the operating point analyzed in the main text ($P_c=60$~mW, $\Delta_c/2\pi=-1.2$~MHz, $\sigma_{v_z}=0.7$~m/s, the effective width of the dressed sub-ensemble, $3\sigma_{v_y}=40$~m/s).}
    \label{fig:robustness}
\end{figure}

Because the coupling frequency is fixed, only a narrow transverse-velocity class is dressed, consistent with Figure~\ref{Fig:fitting}. All efficiencies therefore refer to the molecules that genuinely interact with the coupling field. Reference ensemble: the value \(\sigma_{v_z}=0.7\)~m/s used below is the effective transverse-velocity width of the dressed sub-ensemble, that is, of the molecules that the fixed-frequency coupling field engages, and not the width of the beam as a whole; the latter follows from the \(6.77\)~MHz Doppler FWHM of the coupling-free spectrum as \(\sigma_{v_z}\approx12.6\)~m/s (Section~S2). Because the selection acts only along the laser axis, the axial width \(\sigma_{v_y}\) is unaffected and keeps the measured beam value. All populations and robustness maps here are therefore quoted per dressed molecule, and the useful yield additionally carries the dressed fraction, about one fifth of the beam at the present collimation. Figure~\ref{fig:robustness} reports the dependence of $P_{\mathrm{final}}$ on the coupling power $P_{\rm c}$, the coupling detuning $\Delta_{\rm c}$, and the velocity widths $\sigma_{v_y}$ and $\sigma_{v_z}$, with $\Delta_c<0$ and the followed branch $E_{+,-1}$.

A high final population is obtained over a broad range of $P_{\rm c}$ and $\Delta_{\rm c}$ [Figure~\ref{fig:robustness}(a1)]. At $\Delta_c/2\pi=-1.2$~MHz, $P_{\mathrm{final}}$ increases with $P_{\rm c}$ and then reaches a broad plateau, $P_{\mathrm{final}}\gtrsim0.85$ for $P_{\rm c}\approx50$--120~mW [Figure~\ref{fig:robustness}(b1)]; at fixed $P_{\rm c}$ a broad high-population region spans $\Delta_c/2\pi\approx-3$ to $-1$~MHz [Figure~\ref{fig:robustness}(c1)]. The scheme thus does not rely on a finely tuned operating point. The velocity-width dependence shows the greater sensitivity in the transverse direction: $P_{\mathrm{final}}$ degrades appreciably beyond $\sigma_{v_z}\approx2$~m/s of dressed-class width and falls to about 0.5 near 4~m/s, while it varies by only a few percent across $\sigma_{v_y}\leq120$~m/s [Figure~\ref{fig:robustness}(a2)--(c2)]. Control of the transverse velocity distribution is therefore the more critical practical requirement.
